\documentclass[10pt,journal,compsoc]{IEEEtran}

\usepackage{graphicx}
\usepackage{multicol}
\usepackage{multirow}
\usepackage{url}
\usepackage{xcolor}
\usepackage{amsmath}
\usepackage{verbatim} 
\usepackage{csquotes}
\usepackage{booktabs}
\usepackage[lofdepth,lotdepth]{subfig}
\usepackage{tikz}
\usetikzlibrary{matrix,calc}
\newtheorem{definition}{Definition}
\usepackage{algorithm} 
\usepackage{algpseudocode} 
\usepackage{ragged2e}
\usepackage{float}

\begin{document}
	\title{Visualization of Unstructured Sports Data - An Example of Cricket Short Text Commentary}
	
		\author{Swarup~Ranjan~Behera%
			~and~Vijaya~V~Saradhi
			\IEEEcompsocitemizethanks{\IEEEcompsocthanksitem S. R. Behera and V. V. Saradhi are with the Department of Computer Science and Engineering, Indian Institute of Technology, Guwahati,
				India, 781039. \protect E-mail: \{b.swarup, saradhi\} @ iitg.ac.in}
}

	\markboth{INFOVIS 2024 PREPRINT}%
	{Shell \MakeLowercase{\textit{et al.}}: Bare Demo of IEEEtran.cls for Computer Society Journals}
	
	\IEEEtitleabstractindextext{%
		\justify
		\begin{abstract}
Sports visualization focuses on the use of structured data, such as box-score data and tracking data.  Unstructured data sources pertaining to sports are available in various places such as blogs, social media posts, and online news articles. Sports visualization methods either not fully exploited the information present in these sources or the proposed visualizations through the use of these sources did not augment to the body of sports visualization methods.  We propose the use of unstructured data, namely \textit{cricket short text commentary} for visualization.  The short text commentary data is used for constructing individual player's strength rules and weakness rules. A \textit{computationally feasible definition} for player's strength rule and weakness rule is proposed. A visualization method for the constructed rules is presented. In addition, players having similar strength rules or weakness rules is computed and visualized.  We demonstrate the usefulness of short text commentary in visualization by analyzing the strengths and weaknesses of cricket players using more than one million text commentaries. We validate the constructed rules through two validation methods. The collected data, source code, and obtained results on more than 500 players are made publicly available.
		\end{abstract}
		
		\begin{IEEEkeywords}
			Cricket short text commentary, strength and weakness visualization, sports data visualization.
	\end{IEEEkeywords}}
	
	\maketitle
	
	\IEEEdisplaynontitleabstractindextext
	\IEEEpeerreviewmaketitle
	
	\IEEEraisesectionheading{\section{Introduction}\label{sec:introduction}}

	\IEEEPARstart{U}{nstructured text} data is witnessing substantial growth every  year\footnote{\url{https://bit.ly/340WWZV}}.  The sources of unstructured text data include blogs, microblogs, news articles, and social media posts. Utility of this data is found in diverse applications such as information retrieval~\cite{Mooney}, clustering~\cite{Beil}, classification~\cite{Sriram}, and event detection~\cite{earth}. Visualization methods have also given importance to the unstructured text data~\cite{textvis} and are employed for tasks such as text summarization~\cite{Kuo, Yatani}, sentiment analysis~\cite{Hao}, trend analysis~\cite{Liu}, and pattern analysis~\cite{Nguyen}. 
	
	The growth of unstructured text data has percolated to the sports domain as well~\cite{sportsviz}. The existing sports visualization methods make use of the unstructured text available in the form of online social media and news articles. Wongsuphasawat~\cite{WONGSUPHASAWAT} proposed a visualization approach that uses the tweets to highlight the exciting moments during soccer games. The idea in this work is that exciting events such as shots or missed shots reflect the peaks on twitter. Similarly, Marcus et al.~\cite{Twitinfo} proposed TwitInfo to visualize and summarize the peak soccer events on twitter. These efforts centered around the identification of important events during a game. However, the identified events through the above methods can also be identified using structured data such as box-score data and tracking data. From this perspective, the use of unstructured data in the sports domain did not augment the body of sports visualization. 
    Unstructured sports data are not fully explored and are untapped resources in the sports domain. The unstructured data, such as the authentic ball-by-ball text commentary, is found in the game of  Cricket\footnote{\url{https://www.icc-cricket.com/video/what-is-cricket}}. The present work is an attempt to analyze the \textit{cricket short text commentary} (each text commentary contains a maximum of 50 words). As the nature of structured data (box-score data or tracking data) is significantly different from that of the unstructured data (short text commentary) for a given match, the use cases of cricket short text commentary analysis augment to the body of cricket visualization methods.
	
    Various graphical representations are obtained using the structured data generated from multiple sensors deployed on the cricket field~\cite{Ayan}. These graphs summarize and present the - (i) overall progress of the game along the lines of runs and wickets (manhattan chart and skyline view), (ii) individual player's batting overview (wagon wheel and ground map), (iii) individual player's bowling overview (pitch map), and (iv) fielding overview (field position map). These are used predominantly by broadcasters, team management, match organizers, and audience. These representations capture the game's play on a macroscopic level and do not attend to the minute details. Experts, including team coach and team management, on the other hand, require the knowledge of minute details of the game to improve the performance of individual player and to plan strategies. To obtain microscopic information, experts rely on data generated from ball tracking and player tracking technologies. However, tracking data are limited to specific events of the game and are devoid of the rules and context of the game.
	
    The objective of this work is to propose visualization methods for cricket using unstructured short text commentary data.  In particular, we utilize the minute details present in the short text commentary data to (i) propose a computational model to find the strength and weakness rules of individual player, and (ii) present a set of visualizations for viewing and interpreting the obtained rules. The major contributions of this work are as follows:

	\begin{itemize}
		\item The primary source of data for graphical representation in cricket are box-score data, tracking data, and meta-data. For the first time, the use of unstructured cricket text commentary data is proposed. As the unstructured short text commentary differs significantly from the structured data, the proposed analysis augments the traditional graphical representations in cricket.

		\item Propose to identify the strength and weakness rules of individual player using short text commentary data. As the description of strength and weakness rules are subjective, a \textit{computationally feasible} definition of strength rule and weakness rule is introduced.
		\item A visualization method is proposed for each of the following:
		\begin{itemize}
			\item For interpreting the obtained strength and weakness rules of individual player.
			\item To find players having similar strengths and similar weaknesses.
		\end{itemize}
		\item Validation of the obtained strength and weakness rules is presented.
	\end{itemize}
	
	\section{Related Work}\label{sec:RW}
	Literature closely related to sports data visualization and short text visualization is presented in this section.
	
	\subsection{Sports Data Visualization}\label{sec:SV}	
	Sports data visualization has attracted the attention of the practitioners as well as academics~\cite{Perin1}. Literature based on sports data is classified into three categories - (i) box-score data, (ii) tracking data, and (iii) meta-data.
	
	\textbf{\textit{Box-score Data:}}	
	Box-score data are the discrete data referencing in-game events. It can be the summary statistics or the finer-grained event data. Humans generate most of these events and statistics. For example, the tabular presentation of a soccer teams' points. SportsVis~\cite{SportsVis} uses the baseline bar display and player map to explore baseball team's and player's performance throughout a season, respectively. However, it considers only the aggregate information for any particular game. Buckets~\cite{Peter} utilizes the basketball shot data to view details about a single player, compare multiple players, and explore the league trends. 
	
	\textbf{\textit{Tracking Data:}}
	Recent developments in tracking and sensing technologies enabled the collection of the spatio-temporal information (x and y coordinates at time t) about the players and equipments in real-time during the play. Many visualization tools are introduced for finding the hidden patterns using the spatio-temporal data. In soccer, ForVizor~\cite{ForVizor} visualizes the formation changes over time and reveal the continuous spatial flows of formations. It involves manual annotation of the entire video by experts, which is a significant effort and may not be scalable. Baseball4D~\cite{DIETRICH} makes use of the raw baseball tracking data (player and ball) over time and plot them as events on a dot map to reconstruct the entire game and visually explore each play. Spatio-temporal data based analysis focuses on visualizing low-level information (e.g., player actions). High-level tactical strategies, e.g., team tactics, are hard to infer from this low-level information.
	
	\textbf{\textit{Meta Data:}}
	Meta-data are the additional data surrounding sports that can add context and content details to the box-score and tracking data. Examples of meta-data are the physical characteristics of players, stadium capacities, and social media activity during a game. A challenging problem in sports is to highlight the interesting moments in a game. Wongsuphasawat~\cite{WONGSUPHASAWAT} and Marcus et al.~\cite{Twitinfo} proposed  visualization approaches that use tweets to highlight the exciting moments during soccer games. However, meta-data has no unique agreement on what constitutes meta-data in the context of a game.
	
	Baseball, which is considered as cricket's direct descendant, has significant similarity with cricket. Due to this similarity, many statistics in cricket are inspired by the popular statistics in baseball. However, cricket has a very fewer number of visualization methods with limited utilities\footnote{\url{https://bit.ly/2QFLqPE}}. CricVis~\cite{Ayan} is a web-based visualization system that utilizes the box-score data (scorecards) and the tracking data (ball tracking) to construct visualizations such as pitch maps and stump maps to analyze the bowling and batting overview, respectively. The proposed visualizations in cricket mainly utilize structured data such as box-score data and tracking data. 

	\subsection{Short Text Visualization}\label{sec:STV}
	
    Short texts are usually less than a few hundred characters long. Examples of short texts include microblogs, online chat records, product reviews, and blog comments. Unlike conventional texts (refer to ~\cite{survey,wordle,WordTree,cartographic,InfoSky,SparkClouds,ThemeRiver} for conventional text visualizations), short texts are sparse (contains few words), making the job of information extraction difficult.  To provide a high-level summary of social media text, word clouds~\cite{TagCrowd} are used. SentenTree~\cite{SentenTree} visualizes the frequent sequential patterns in social media text to gain a fast understanding of critical concepts and opinions. It uses a node-link diagram where nodes are words, links are the word co-occurrence within the same sentence, and the size of nodes indicate their frequency of occurrence. Opinions (posts and comments) in social forums are generally organized based on their semantic contents~\cite{Joty}. However, it encourages selective exposure to information and opinion polarization. Gao et al.~\cite{Gao} developed an interface to allow interactive visualization and categorization of Reddit posts about controversial topics that involve people with different attitudes and stances. The aim is to allow users to explore opinions from different stances and ultimately reduce opinion polarisation.
	
	The sports visualization methods present in the literature mainly focus on structured data. To the best of our knowledge, this is the first visualization work that utilizes the unstructured data in the form of short text commentary. Cricket text commentary is a short text with some unique features. None of the above-discussed visualization methods can be employed directly in the context of cricket text commentary as the objective of the present work is distinctly different from the rest of the work discussed above. In particular, the relationship between batting and bowling features, which are collectively present in the short text document, needs to be captured.
    
    \section{Background, Motivation, and Design Goals}\label{sec:BSO}
    This section presents the background of various forms of data in cricket. The motivation for the present work is discussed in the context of unstructured short text commentary. Fundamental requirements are identified, and associated design goals are derived, followed by system overview.
	
	\subsection{Background}\label{sec:B}
	
    In cricket, box-score data (e.g., scorecards) are the first kind of data that is recorded (in the year 1772). 
    Even though it is almost 247 years old, the change in the box-score data percolates only for a few places. Accordingly, the change in the box-score data visualization made progress in this available additional information. 
    The use of advanced data capture techniques\footnote{\url{https://tinyurl.com/y4upvcgf}} in cricket enabled the coaches and broadcasters to uncover new insights into every shot, run, and wicket. Hawk-Eye technology is used for tracking ball trajectory, speed of the delivery, position of the pitched ball, and bounce of the ball. The visualizations associated using this data is employed in decision making by umpires, used by analysts and experts. 
    The main bottleneck with tracking data is that these are not available publicly. In addition, they are highly expensive to deploy in every match. 
    In cricket, meta data mostly account for conditions that are outside the context of the game. Examples include, moisture condition, pitch condition like crack or grass available on the pitch. These are not actively used for visualization purposes.
	
    The box-score data and tracking data are structured data that capture only a limited amount of information specific to the events in matches. Unstructured data such as video broadcast, audio commentary, and text commentary capture all the information for matches and events specific to matches with minute details. In addition to the above three, social media posts and news articles add to the multiple sources of unstructured data. However, text commentary data, a form of short text, has some unique advantages over all of them. As compared to video or audio data, it is readily available and requires less storage and computational resources. Unlike social media posts, it is credible, maintains a consistent style,  and provides the mapping of each delivery to a single text commentary. Unlike news articles, it provides a microscopic view with ball-by-ball details of the match. Even though the text commentary data has been around for quite some time, it still is an untapped and unexplored resource. 
	
	\subsection{Motivation}\label{sec:M}
    For box-score data and tracking data, various visualization techniques are provided. However, for the unstructured data, no such visualization techniques exist. In this work, unstructured data is used for discovering information that is not present in the structured data (box-score data and tracking data).  Aim of this work is to augment the body of the visualization techniques with a set of new visualizations using unstructured data. From this perspective, short text commentary data is analyzed to build player specific strategies in the form of visualization.
    
	\subsection{Requirement Analysis}\label{sec:RA}
    Given the minute details present in the short text commentaries, the following questions are addressed from the visualization perspective.
	
	\textbf{R1}  \hspace{.1cm} \textbf{What are the strength rules and weakness rules of a player? Can these rules be \textit{visualized}?} 
	To provide a \textit{computational definition} of strength rule and weakness rule of individual player. Such a definition is of help and is the foundation of future automated strategy planning for cricket.
		
	\textbf{R2}  \hspace{.1cm} \textbf{Which players have similar strengths and weaknesses?} 
	Grouping similar players based on their strengths and weaknesses also yield in-depth insights. It will help in devising strategies for team member selection.
	
	\subsection{Design Goals}\label{sec:DG}
	Based on the identified requirements, the following two design goals are derived.
	
	\textbf{D1}  \hspace{.1cm} \textbf{Visualization of strength and weakness rules of individual player.} A visualization method is introduced that specifically aims to represent the strength and weakness rules. The proposed method's flexibility is to extract and visualize the strength and weakness rules in variety of scenarios such as (i) strength/weakness of a player against a particular player (ii) strength/weakness of a player against a particular team (iii) strength/weakness of a player in a series etc.
	
    \textbf{D2}  \hspace{.1cm} \textbf{Visualization of similar players based on their strengths and weaknesses.} The design goal D1 is extended to visualize similar players based on their strength rules or weakness rules.
    
    \subsection{System Overview}\label{sec:SO}
	
	\begin{figure}[!t]
		\begin{center}
			\includegraphics[scale=0.97]{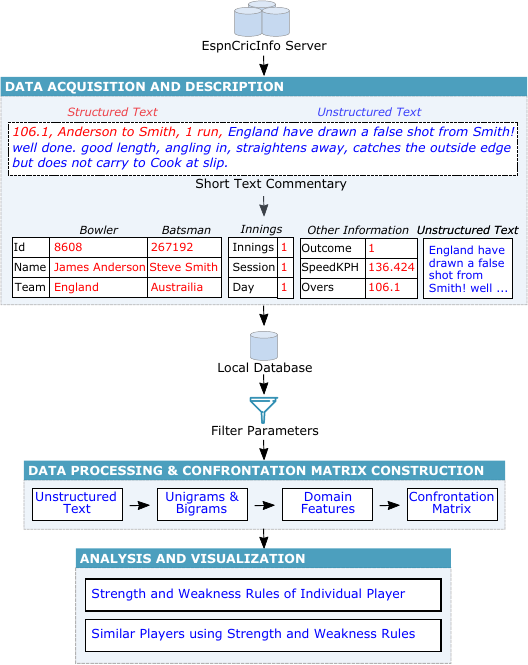}
			\caption{System Overview.}
			\label{fig:overview}
		\end{center}
	\end{figure}
	
	Fig.~\ref{fig:overview} presents the system overview in addressing the design goals given in section 3.4. It has three modules. The \textit{data acquisition and description} module deals with the acquisition and description of short text commentaries. The \textit{data processing and confrontation matrix construction} module contain three components: (i) processing of text commentaries to represent each text commentary as a set of unigrams and bigrams, (ii) mapping of the obtained unigrams and bigrams to a set of domain-specific features, and (iii) representation of the extracted features in a confrontation matrix. The confrontation matrix records the text commentary frequencies for every pair of batting and bowling features. The \textit{analysis and visualization} module contains two visualization components: (i) strengths and weaknesses rules of an individual player and (ii) similar players based on their strengths and weaknesses rules. 
	
	\section{Data Acquisition and Description}\label{sec:DA}
    Cricket has multiple formats of the game, of which, the \textit{Test cricket} format is considered for the present work. EspnCricInfo\footnote{EspnCricInfo (\url{http://www.espncricinfo.com/}) is a sports news website exclusively dedicated to the game of cricket. This website provides the most comprehensive repository of cricket text commentary.} is selected as the data source. The earliest documented text commentary available for the Test matches dates back to 2006. To collect text commentary associated with a given Test match, one has to first obtain the season and series in which this particular match is a part. In addition, match and innings IDs and associated URLs need to be formulated from ESPNCricInfo's archive.  This information is used to acquire the short text commentaries for a given match. This procedure is repeated for all the matches played between May 2006 to April 2019. Total text commentaries of 1,088,570 deliveries are collected spanning thirteen years and account for a total of 550 international Test cricket matches. 
    The acquired short text commentaries are stored in a local database. 

    Consider the example of short text commentary presented in Fig.~\ref{fig:overview}. This commentary describes the first delivery in the 107th over of the game in which \textit{Anderson} is the bowler and \textit{Smith} is the batsman. The outcome of this delivery is one run. The rest of the text describes the way the ball is delivered, and the way batsman played it. For instance, this commentary describes several features of bowling, such as length (\textit{good length}) and movement (\textit{angling in}). Similarly, it describes batting features such as response (\textit{outside edge}) pointing to the batsman's imperfection. The word \textit{false shot} emphasizes the imperfection and points to batsman's weakness against \textit{good length} and \textit{angling in} delivery. A large number of deliveries played by batsman Smith are analyzed. When consistent imperfection is observed on deliveries with similar features, it leads to weakness of Smith. Such detailed weakness rules are far more expressive and useful than simple statistics such as batting averages.    

	\section{Data Processing and Confrontation Matrix Construction}\label{sec:DPFE}
	
		\begin{table}[bt]
		\begin{center}
			\caption{List of Words Relevant in Cricket but Considered as Stop Words in Other Application Domains.}
			\resizebox{1\columnwidth}{!}{%
				\begin{tabular}{|l|l|l|l|l|l|l|l|} \hline
					off & on & room & across & behind & back & out \\
					\hline
					good & great & into & away & up & down & long \\
					\hline
					turn & point & from & further & under & full & open \\ \hline
				\end{tabular}
			}
			\label{tab:stop}
		\end{center}
	\end{table}
	
	\begin{table}[bt]
		\begin{center}
			\caption{Feature Definition of
				\textit{Beaten} ($FD_{beaten}$)
				.}
			\resizebox{1\columnwidth}{!}{%
				\begin{tabular}{|l|l|l|l|l|} \hline
					miss & deceive & poor shot & doesn't time & can't connect \\
					\hline
					beat & weak & defeat & clue less & knock down \\
					\hline
					edge & trap  & top edge & inside edge & bottom edge \\
					\hline
					confuse & fumble & lucky & miss hit & miss judge \\
					\hline
				\end{tabular}
			}
			\label{tab:beaten}
		\end{center}
	\end{table}
	
		
	\begin{table*}[bt]
		\centering
		\caption{Batting and Bowling Features}
		\resizebox{\textwidth}{!}{%
			\begin{tabular}{p{10cm} p{6cm}}
				\toprule
				\textbf{Features}& \textbf{Feature Description}\\
				\midrule
				\multicolumn{2}{l}{ \textit{\textbf{Batting features (Outcome, Response, Footwork, and Shot area)}}} \\ 
				\midrule
				\textbf{0, 1, 2, 3, 4, 5, 6( also 6$+$)} runs and \textbf{out}  & Outcome of a particular delivery \\
				\textbf{Beaten} (exhibits imperfection), \textbf{Defended} (blocks or leaves the ball), \textbf{Attacked} (plays aggressive shots) & Response of the batsman on each delivery \\
				\textbf{Front foot} (ball is played in front of the batsman), \textbf{Back foot} (played behind the batsman's wicket) & Footwork of a batsman when facing a delivery \\
				\textbf{Third man, Square off, Long off, Long on, Square leg, Fine leg} & Region where shot is played by batsman \\
				\midrule
				\multicolumn{2}{l}{ \textit{\textbf{Bowling features (Length, Line, Type, Speed, and Movement)}}} \\ 
				\midrule
				\textbf{Short} (closer to the bowler), \textbf{Good} (optimal length, in between short and full), \textbf{Full} (nearer the batsman)  & How far down the pitch (length) the ball bounces\\
				\textbf{Off} (on or outside off-stump), \textbf{Middle} (on middle-stump), \textbf{Leg} (on or outside leg-stump) & How far to the left or right (line) of the wicket ball is travelling \\
				\textbf{Spin} (slow deliveries which turn sharply after pitching), \textbf{Swing} (fast deliveries with movement in the air) & Nature (type) of the delivery \\
				\textbf{Fast} (medium: 60-80 mph, fast: 80+ mph), \textbf{Slow} (40-60 mph) &  Speed of the ball after it is released \\
				\textbf{Move-in} (towards batsman), \textbf{Move-away} (away from batsman) & Movement of the ball \\
				\bottomrule
		\end{tabular}}
		\label{tab:features}
	\end{table*}

	Domain-specific technical vocabulary is dominantly observed in the unstructured part of short text commentary. However, this technical vocabulary has two fundamental challenges:
	\begin{itemize}
		\item \textbf{Stop words:} Majority of this vocabulary refers to stop words in the conventional information retrieval literature. The non-exhaustive list of technical stop words is presented in table~\ref{tab:stop}.
		\item \textbf{Sparsity:} The vocabulary size is large, and the unstructured component in one delivery of cricket commentary has only up to 50 words.  
	\end{itemize}
	To overcome these two challenges, domain-specific batting features and bowling features are identified.  Each feature is represented as a \textit{set} of unigrams and bigrams such that the identified set corresponds to the feature in question and is represented as feature definition FD.  As an example, for the batting feature \textit{beaten}, the corresponding unigrams and bigrams are as given in table~\ref{tab:beaten}. This unigram/bigram to feature mapping is obtained by consulting the cricket experts. A total of 19 batting features and 12 bowling features are identified as given in table~\ref{tab:features}.  Corresponding to these features, 19 (batting features) and 12 (bowling features) sets of unigrams and bigrams are obtained. This method of obtaining features has addressed the stop word related problem. The sparsity is addressed by mapping every unigram and bigram to one of the features.
	
    Section~\ref{sec:DA} detailed the process of acquiring the entire commentary database. To achieve the design goals, one has to obtain a \textit{subset of text commentary} from this database. Extraction of this subset depends on a \textit{filter tuple} $\langle$ \textit{player}, \textit{opponent player}, \textit{time}, \textit{type} $\rangle$ having four elements:
    
    \begin{itemize}
    \item \textbf{Player:} The one player for whom strength rules and weakness rules are to be extracted.
	\item \textbf{Opponent Player:}  The opponent player (for a batsman, opponents are the bowlers and vice-versa) can be one player or a set of players.
    \item \textbf{Time:}  The element \textit{time} can be per session, per day, per innings, per match, per series, or an entire career. 
    \item \textbf{Type:}  It can be batting or bowling, depending on whether we want to analyze the batting or bowling of the player in focus. If the type is batting (or bowling), then all the commentaries where the player is mentioned as a batsman (or bowler) are selected. 
	\end{itemize}

	\begin{figure*}[bt]
		\begin{center}
			\includegraphics[scale=1]{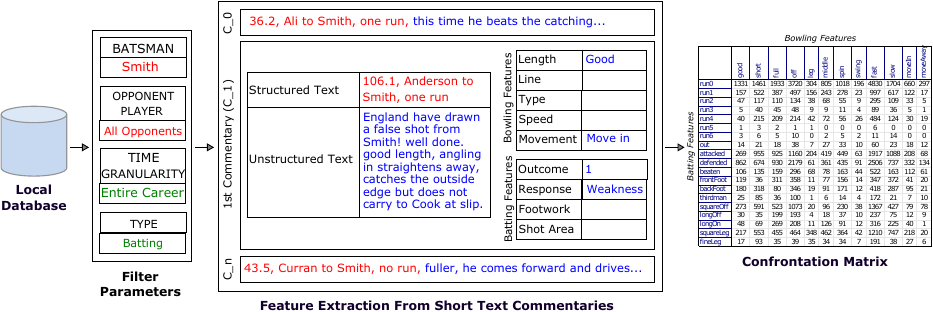}
			\caption{Steps of Confrontation Matrix Construction.}
			\label{fig:CM}
		\end{center}
	\end{figure*}
 
    For achieving the design goal D1, the filter tuple is $\langle$ \textit{Steve Smith}, \textit{all opponents}, \textit{entire career}, \textit{batting} $\rangle$. A subset of the text commentary is extracted based on this filter tuple. From this subset, a confrontation matrix of size $19 \times 12$ is constructed (refer to Algorithm~\ref{algo}) in which rows correspond to batting features of the player and columns correspond to bowling features of opponent players. Fig.~\ref{fig:CM} presents the steps involved in the construction of confrontation matrix for the batsman Steve Smith. Every element in this matrix corresponds to how the batsman confronted with the bowlers. For example, how many numbers of times batsman has shown aggression (attacked) on short length deliveries? The first entry in this matrix is 1331, which indicates that Steve Smith has  scored zero run in 1331 good length deliveries in his whole career (up to April 2019). Similarly, other entries in the confrontation matrix represent the count of co-occurrences of batting and bowling features. Confrontation matrices of all 264 batsmen and 264 bowlers are provided at the link:~\url{https://bit.ly/2QZjx5x}

	\begin{algorithm}
		\caption{Construction of Confrontation Matrix}\label{algo}
		\begin{algorithmic}[1]
			\Require
			Filter tuple, Feature definitions (FD), BattingFeatures, BowlingFeatures, A matrix  $CM_{19\times12}$ of zeros.
			\State Extract the text commentaries using \textit{filter tuple}
			\For {Every commentary}
			\State Initialize two sets: $\underline{bat} = \phi$ and $\underline{bowl}= \phi$
			\State Get the \underline{outcome} (denoted as o) from the structured part of commentary
			\For {i in \{0, 1, 2, 3, 4, 5, 6, out\}}
			\If{o == i}
			\State $\underline{bat} = \underline{bat} \cup \{i\}$
			\EndIf
			\EndFor
			\State Get all the unigrams and bigrams from the unstructured part of commentary
			\For {Every unigram or bigram y}
			\For {Every i $\in$ BattingFeatures}
			\If{y $\in$ $FD_{i}$}
			\State $\underline{bat} = \underline{bat} \cup \{i\}$
			\EndIf
			\EndFor
			\For {Every j $\in$ BowlingFeatures}
			\If{y $\in$ $FD_{j}$}
			\State $\underline{bowl} = \underline{bowl} \cup \{j\}$
			\EndIf
			\EndFor
			\EndFor
			\For {Every $a \in \underline{bat}$ and $b \in \underline{bowl}$}
			\State $CM[a,b] = CM[a,b] + 1$
			\EndFor
			\EndFor
			\State \Return Confrontation Matrix (CM)
		\end{algorithmic} 
	\end{algorithm}

	\section{Visual Design}\label{sec:VD}
    Previous sections detailed two of the three modules, as depicted in Fig.~\ref{fig:overview}. In this section, the third module, namely, analysis and visualization, is presented. In Section~\ref{sec:VSW}, design goal D1 is achieved in two steps; a computational approach for strength rule and weakness rule construction of individual player is first presented, followed by proposing a visualization method through biplots.  Visualization of players having similar strength rules and similar weakness rules (design goal D2) is presented in Section~\ref{sec:VSP}.
	
	\subsection{Visualization of Strength Rules and Weakness Rules of Individual Player}\label{sec:VSW}
    A two-step method is adopted for visualizing strength rules and weakness rules for an individual player - (i) computation of strength rules and weakness rules and (ii) visualization of obtained strength rules and weakness rules.
%
	
	\subsubsection{Computation of Strength Rules and Weakness Rules}\label{sec:DRCA}
	The input for computing the strength rules and weakness rules is the confrontation matrix (detailed in section~\ref{sec:DPFE}), in which both the row as well as the column features are discrete random variables. In order to compute the strength rules or weakness rules, arriving at computationally feasible definition of what constitutes a rule is very important. The following definitions offer the computational ability of strength rule and weakness rule.
	
	\begin{definition}{\textit{\textbf{Rule.}}} \label{d1}
		A rule must comprise of one batting feature and one bowling feature which are \textit{dependent} on each other. 
	\end{definition}

	The dependency is captured through the extent of violation ($\alpha$) of the \textit{independence of events} probability axiom: P(batting feature $\cap$ bowling feature) $= \alpha$ P(batting feature) $\times$ P(bowling feature); where $\alpha = 1$ when the batting feature is independent of the bowling feature. When $\alpha < 1$, then there is a dependency of batting feature on bowling features. The goal here is to obtain the relationships among the batting features and bowling features. Note, however that both of these features are in high dimensional space. Dimensionality reduction methods are traditionally employed for reducing the number of features, capturing the discriminative variables, and for visualization. As the batting and bowling features are discrete random variables, we use Correspondence Analysis (CA)~\cite{doi:10.1177/096228029200100106,beh2014correspondence,GreenacreCB}, a multivariate statistical technique, to obtain a low dimensional subspace that contains these features. To obtain the low dimensional space, CA minimizes the sum of the squared distance from the reduced subspace to each of the feature points. To obtain the solution to this minimization, singular value decomposition (SVD)~\cite{meyer2000matrix} is applied on the normalized and centered confrontation matrix $N$ to obtain the principal components of batting features and bowling features. The complete CA algorithm is presented in Algorithm~\ref{algo1}.
		
		\begin{algorithm}
		\caption{Correspondence Analysis} \label{algo1}
		\begin{algorithmic}[1]
			\Require A confrontation matrix $N_{I\times J}$
			\State Matrix sum: $ n = \sum_{i=1}^{I} \sum_{j=1}^{J} N_{ij}$
			\State Row masses(r): $r_i= \frac{N_{i.}}{n},  i = 1,2, \cdots, I$
			\State Diagonal matrix: $D_r = diag(r_1, r_2,...,r_I)$
			\State Column masses(c): $c_j= \frac{N_{.j}}{n},  j = 1,2, \cdots, J$
			\State Diagonal matrix: $D_c = diag(c_1, c_2,...,c_J)$
			\State Correspondence matrix: $ P = \frac{1}{n}N $
			\State Standardized residuals: $A = D_r^{-\frac{1}{2}} (P - rc^T) D_c^{-\frac{1}{2}}$
			\State Singular value decomposition: $ A = U \Sigma V^T $
			\State Standard coordinates of rows: $ \Phi = D_r^{-\frac{1}{2}}U $
			\State Standard coordinates of columns: $ \Gamma = D_c^{-\frac{1}{2}}V$
			\State Principal coordinates of rows: $ F = \Phi \Sigma $
			\State Principal coordinates of columns: $ G = \Gamma \Sigma $  
			\State \Return F and G
		\end{algorithmic} 
	\end{algorithm} 
	
	Applying CA on the confrontation matrix yields the row (batting) principal components and column (bowling) principal components (F and G, respectively). Note that F retains the batting features of the player, and G retains the bowling features of the opponent players as detailed in Algorithm~\ref{algo1}. A finer interpretation of the relationship between batting features and bowling features for strength rule and weakness rule construction is described below using \textit{Definitions (2) - (5)}.
	
	\begin{definition}{\textit{\textbf{Strength Rule of Batsman.}}}
		In \textit{Definition~(\ref{d1})}, when the batting feature corresponds to \textit{attack} and involves any of the bowling features.
	\end{definition}

\begin{figure}[bt]
		\begin{center}
			\centering
			\subfloat[Subfigure 1 list of figures text][Attacked Inner Product.]{
				\includegraphics[scale=0.45]{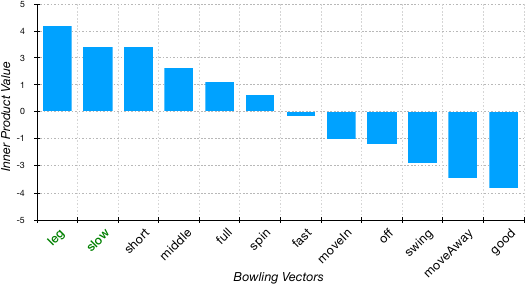}
				\label{fig:InnerA}} 
			\subfloat[Subfigure 2 list of figures text][Beaten Inner Product.]{
				\includegraphics[scale=0.45]{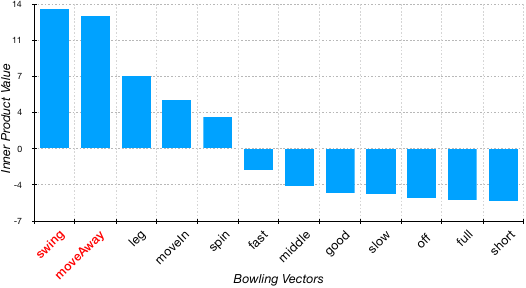}
				\label{fig:InnerB}}
			\caption{Inner Products between $F_{attacked}$ or $F_{beaten}$ and \\ all Bowling Vectors.}
			\label{fig:Inner}
		\end{center}
	\end{figure}

	For a strength rule, the batting feature must have \textit{attacked} feature and the bowling features are identified by taking the inner product of $F_{attacked}$ with every bowling vector of $G$; that is, compute $\langle F_{attacked}, G_j\rangle$ for all $j$. The bowling vector ($G_j$), which yields the highest (or second highest) inner product, corresponds to the first (or second) strength rule of the batsman. 
	The inner products between $F_{attacked}$ and all bowling vectors for batsman Steve Smith are presented in Fig.~\ref{fig:InnerA}. The top two bowling vectors having the highest inner product value with $F_{attacked}$ vector in the decreasing order are  $G_{leg}$ and $G_{slow}$. Thus, the proposed algorithm obtains the first two strength rules of Steve Smith as - (i) ``Smith attacks the deliveries that are bowled on the leg stump'' and (ii) ``Smith attacks the slow deliveries''.
	
	\begin{definition}{\textit{\textbf{Weakness Rule of Batsman.}}}
		In \textit{Definition~(\ref{d1})}, when the batting feature corresponds to \textit{beaten} and involves any of the bowling features.
	\end{definition}

	For a weakness rule, the batting feature must have \textit{beaten} feature and the bowling features are identified by taking the inner product of $F_{beaten}$ with every bowling vector of $G$; that is, compute $\langle F_{beaten}, G_{j} \rangle$ for all $j$. 
	The inner products between $F_{beaten}$ and all bowling vectors for batsman Steve Smith are presented in Fig.~\ref{fig:InnerB}. The top two bowling vectors having the highest inner product value with $F_{beaten}$ vector in the decreasing order are  $G_{swing}$ and $G_{moveAway}$. Therefore, we derive the first two weakness rules for Steve Smith as - (i) ``Smith gets beaten on the deliveries that are swinging'' and (ii) ``Smith gets beaten on the deliveries that are moving away from him''.
	
	Due to the apparent similarity, we are not elaborating on the strength and weakness analysis of the bowlers. However, following are the definitions of strength rule or weakness rule of bowlers.
	
	\begin{definition}{\textit{\textbf{Strength Rule of Bowler.}}}
		In \textit{Definition~(\ref{d1})}, when the opponent's batting feature corresponds to \textit{beaten} and involves any of the bowling features.
	\end{definition}
	
	\begin{definition}{\textit{\textbf{Weakness Rule of Bowler.}}}
		In \textit{Definition~(\ref{d1})}, when the opponent's batting feature corresponds to \textit{attack} and involves any of the bowling features.
	\end{definition}

    The maximum number of rules that can be obtained for one player using the proposed method is 228 (19 * 12). Out of these rules, 12 strength rules and 12 weakness rules are subject to evaluation. Out of these, four dominant rules having a high degree of dependency are presented above. Strength rules and weakness rules corresponding to all the 264 batsmen and 264 bowlers are provided at the link: https://bit.ly/2QZjx5x

	\subsubsection{Visualization of Strength Rules and Weakness Rules}\label{sec:VSWIP}
	
    In order to visually interpret the relationship between batting features and bowling features, the first two principal directions of F and G are obtained from algorithm~\ref{algo1} and plot them on a two-dimensional plot. This method is well known as biplot~\cite{10.2307/2334381}. In the biplot, the row (batting) and column (bowling) vectors having the highest inner product values are the closest vectors. These two vectors constitute a strength rule or a weakness rule.
	
\begin{figure}[bt]
		\begin{center}
			\includegraphics[width=0.9\columnwidth]{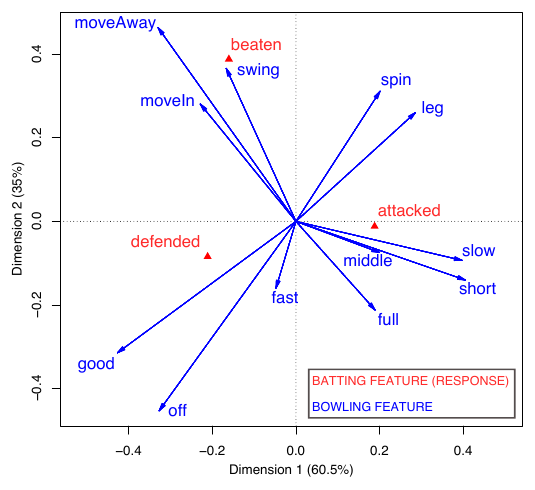}
			\caption{Smith's Response on Various Deliveries.}
			\label{fig:smith-response}
		\end{center}
	\end{figure}

    Fig.~\ref{fig:smith-response} presents the contribution biplot~\cite{GreenacreCB}, a variant of biplot described in~\cite{10.2307/2334381}, which is the biplot depicting the strengths and weaknesses of batsman Steve Smith. For better visualization, instead of having all the batting features in one plot, only a subset of batting features (subset correspondence analysis~\cite{GreenacreCB}) and all the bowling features are presented. Based on his response, (i) the top two closest vectors for $F_{attacked}$ vector in the decreasing order are $G_{leg}$ and $G_{slow}$ and (ii) the top two closest vectors for $F_{beaten}$ vector in the decreasing order are $G_{swing}$ and $G_{moveAway}$. These rules from the biplot are identical to the one given in the computational method presented above. Biplots of all 264 batsmen and 264 bowlers are provided at the link: https://bit.ly/2QZjx5x

\subsection{Visualization of Similar Players Based on Their Strength Rules and Weakness Rules}\label{sec:VSP}
	
	
    In order to realize the design goal D3, confrontation matrices of all batsmen (or all bowlers) are obtained using a fixed filter tuple \textit{$\langle$player, all opponents, entire career, batting (or bowling)$\rangle$}. Each of these matrices is subject to CA, and the first strength rule and first weakness rule of each batsman/bowler is obtained through the computational procedure detailed in section~\ref{sec:VSW}. That is, obtain $F_i, G_j$ and concatenate these two vectors as $(F_i, G_j)$ for each player. These vectors lie in 31 (19 + 12) dimension space representing each player's strength or each player's weakness. To visualize these high dimensional vectors, t-SNE algorithm~\cite{tsne} is employed, which obtains a two-dimensional plot in which players having similar strengths (or similar weaknesses) are placed together as clusters. The analysis of similar batsmen based on their strength and weakness rules are presented below.
    

    \begin{figure}[bt]
		\begin{center}
			\includegraphics[scale=1]{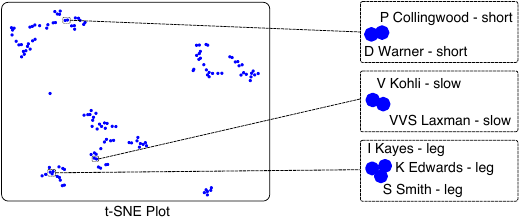}
			\caption{Visualization of Similar Batsmen based on their Strength Rule using t-SNE Plot.}
			\label{fig:Strengthtsne}
		\end{center}
	\end{figure}
	
    Fig.~\ref{fig:Strengthtsne} is the t-SNE plot of all the batsmen, based on their strength vectors $(F_{attacked}, G_j)$. Each point in the plot is the strength vector of a player. Each of these points is presented in the form of \textit{[player name - bowling feature]}, i.e.,  the player has attacked or exhibited strength in the deliveries having this bowling feature. Intriguing player clusters are readily apparent, with batsmen attacking \textit{short length} and \textit{leg line} deliveries in the upper left and lower left corner, respectively. We have considered a total of 264 batsmen from 12 countries. The three squares on the right-hand side of the t-SNE plot show the zoomed versions in which similarities of batsmen are clearly observed. On the top square, batsmen Warner and Collingwood exhibits strength on the short-pitched balls. On the bottom square, Steve Smith has similarity with two batsmen (Kayes \& Edwards), i.e., they exhibit strength when the balls are pitched towards leg stump. 
	
    \begin{figure}[bt]
		\begin{center}
			\includegraphics[scale=1]{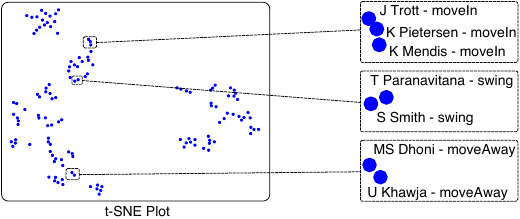}
			\caption{Visualization of Similar Batsman based on their Weakness Rule using t-SNE Plot.}
			\label{fig:Weaknesstsne}
		\end{center}
	\end{figure}
	
	Similarly, Fig.~\ref{fig:Weaknesstsne} is the t-SNE plot of all the batsmen, based on their weakness vectors $(F_{beaten}, G_j)$. Each point in the plot, is weakness vector of a player, presented in the form of \textit{[player name - bowling feature]}. Intriguing player clusters are readily apparent, with batsmen got beaten on \textit{move away} deliveries in the lower-left corner. The three squares on the right-hand side of the t-SNE plot show the zoomed versions in which similarities of batsmen are clearly observed. On the top square, the batsmen exhibit weakness on deliveries, which are moving in towards batsman. In the middle square, Steve Smith has similarity with Paranavitana, i.e. they exhibit weakness on the swinging deliveries.
	
	\section{Validation}\label{sec:V}
	
	\begin{figure*}[!htbp]
		\begin{center}
			\centering
			\subfloat[Subfigure 1 list of figures text][Outcome.]{
				\includegraphics[scale=0.33,angle=270]{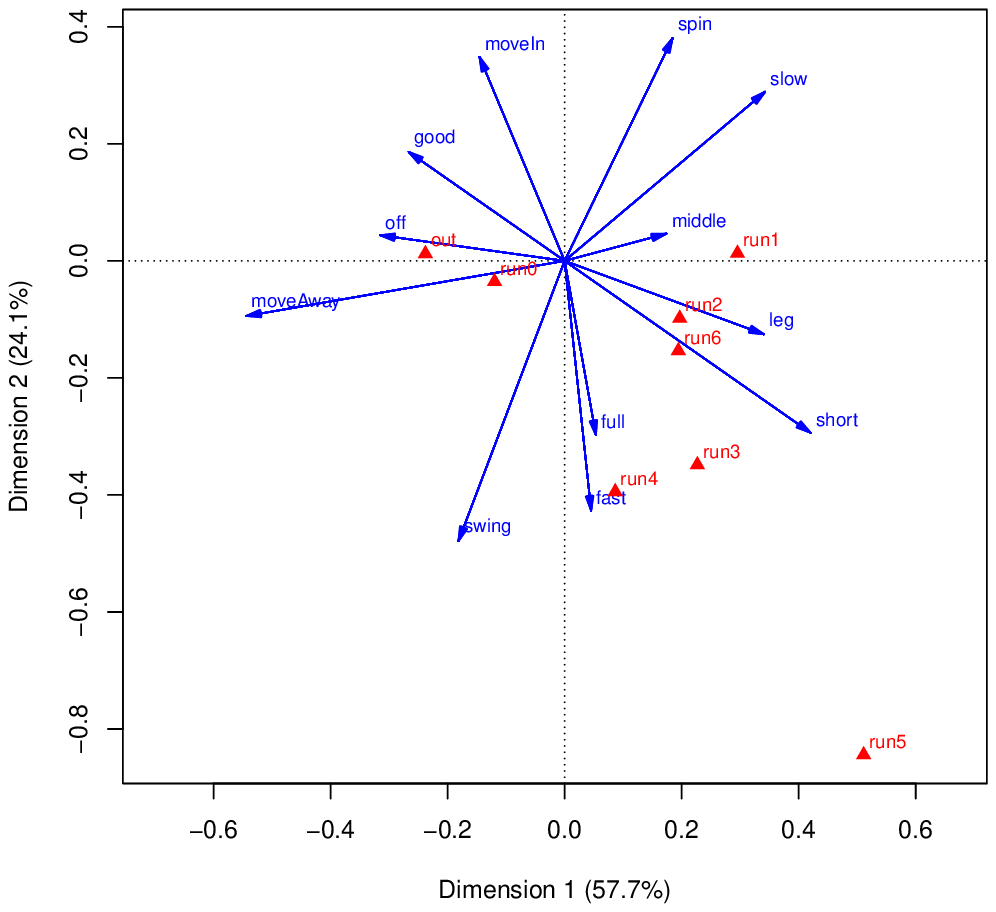}
				\label{fig:InnerA1}} 
			\subfloat[Subfigure 2 list of figures text][Shotarea.]{
				\includegraphics[scale=0.33,angle=270]{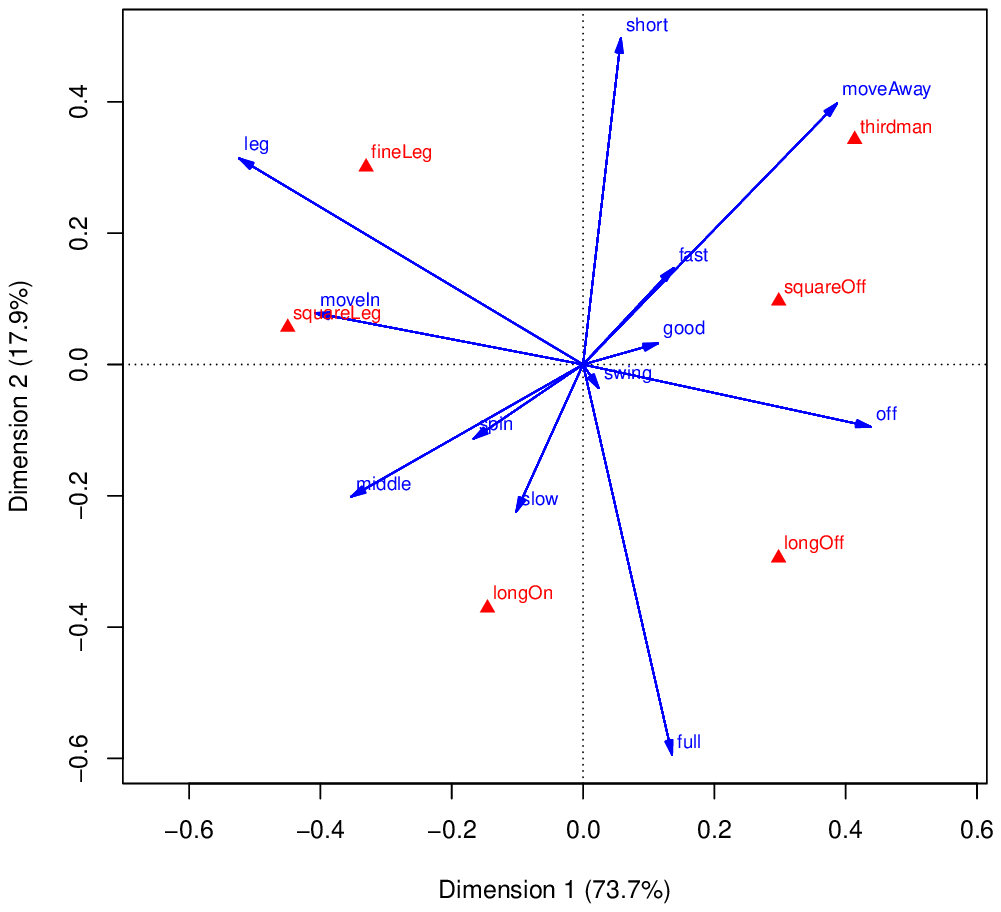}
				\label{fig:InnerB1}}
			\subfloat[Subfigure 3 list of figures text][Footwork.]{
				\includegraphics[scale=0.33,angle=270]{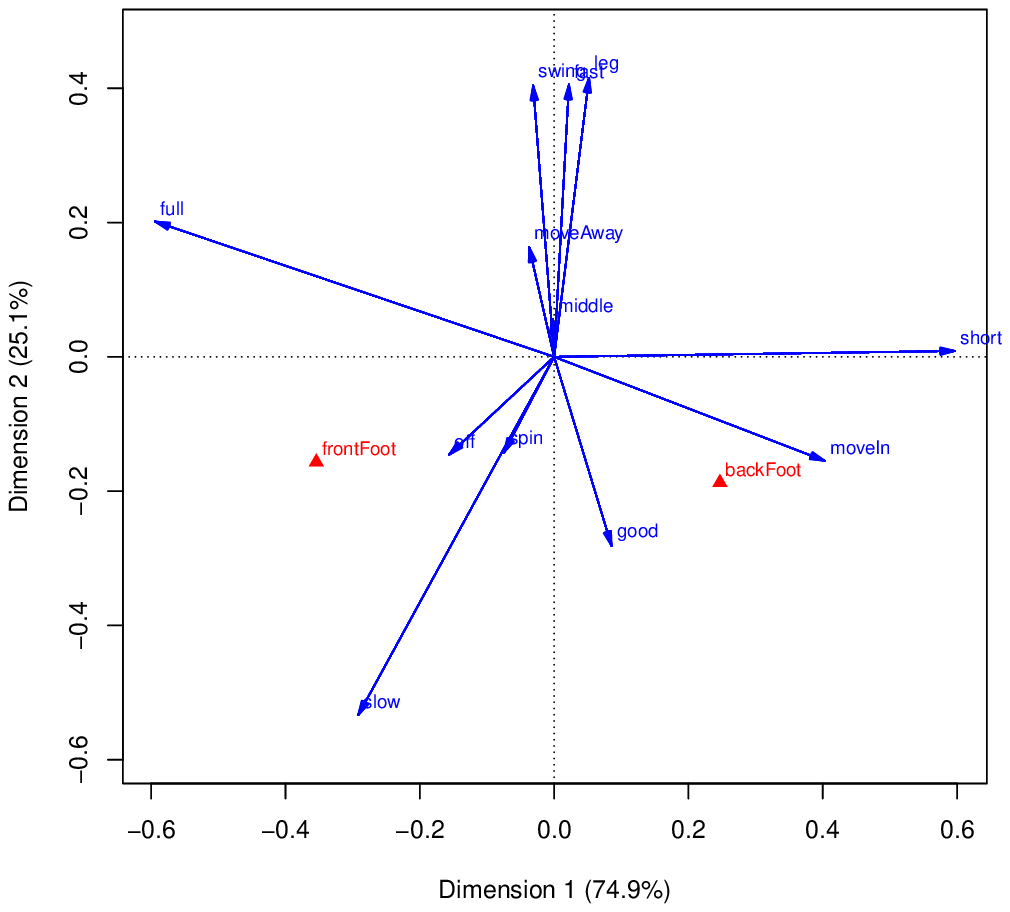}
				\label{fig:InnerC1}}
			\caption{Smith's Outcome, Shotarea, and Footwork on Various Deliveries.}
			\label{fig:Inner1}
		\end{center}
	\end{figure*}

    
    Two distinct validation methods are proposed for D1. (i) For a particular player, namely Steve Smith, ground truth rules through expert analysis are extracted. Obtained rules through CA method for Steve Smith are compared against the expert's ground truth. (ii) As performing expert validation for all the players is a predominantly difficult task, a statistical test, namely \textit{Procrustes Analysis} is employed, which takes two biplots and scores them based on their similarity.

	\subsection{Expert Analysis}
    For expert analysis, the obtained rules through CA method are verified against expert sources. The main bottleneck is the absence of trustable gold standard data about the strengths and weaknesses of every cricket player. One video blog from the domain expert (\textit{Sanjay Manjrekar}) in which the expert shares the strength and weakness rules of Steve Smith is identified, and its analysis is presented below. He has published a video\footnote{\url{https://es.pn/2s0nM55}} with ESPNCricInfo titled -``\textit{What is Steve Smith's weakness?}" on  $1^{st}$ June, $2017$. In this, the author has given one strength rule and one weakness rule. The strength and weakness rules are extracted from the video. These rules are then compared with the ones obtained from the CA-based method. 
    
    Following is a transcript of selective parts of the video along with the rules obtained by the CA method. Expert strength rule (video time $0.15^{''}$): ``Bowlers think he is very vulnerable to the ball pitching on the leg stump, ..., Bowlers tend to attack him on the stump (leg or middle stump line), but then his wonderful angle of the bat carves everything on the leg side. Gets a lot of runs on the leg side''. That is, Steve Smith scores a lot of runs on the leg side for balls bowled on the leg/middle stump. Obtained/CA strength rule: ``Steve Smith attacks deliveries that are bowled on the leg stump''. Expert weakness rule (video time: $0.49^{''}$): ``Bowl seamers as much as possible.'' Obtained/CA weakness rule: ``Steve Smith gets beaten on the move away (seam) deliveries.'' Both the obtained rule through CA is agreeing with the expert constructed rules.
	
	\subsection{Procrustes Analysis}
    The Procrustes analysis~\cite{oro2736, Neto} is used to compare two biplots. The main idea is to minimize the sum-of-squared differences between the two biplots. This test performs the least square superimposition of one biplot to another reference biplot whose output is the sum of squared residual. The lower is this sum the greater is the similarity between the two biplots.
%
%
%
%
%
%
%
%
%
%

	\begin{table}[tb]
		\centering
		\caption{Procrustes Analysis}\label{tab:intrinsic batsman}
		\begin{tabular}{l c c c}
			\toprule
			\textbf{Batsman}   &  $\#Balls_{train}$  & $\#Balls_{test}$   & $ \Delta_{12}^2 $\\
			\midrule
			Joe Root   &  8376  & 2421   & 0.09 \\ 
			Dimuth Karunaratne   &  4898  & 1831    & 0.11 \\ 
			Steve Smith   &  9250  & 1948   & 0.17 \\ 
			Cheteshwar Pujara   &  7947  & 1546    & 0.27 \\ 
			Dean Elgar   &  4475  & 2597    & 0.28 \\ 
			Virat Kohli   &  8085  & 1482   & 0.30 \\ 
			David Warner   &  6999  & 1557    & 0.47 \\ 
			Kane Williamson   &  10165  & 439   & 0.47 \\ 
			\bottomrule
		\end{tabular} 
	\end{table}
	
	For conducting the procrustes analysis in the present setting, standard \textit{data holdout strategy} is employed in which the commentary data after applying the filter tuple to the entire commentary database is divided into training and test data sets. In particular, last year of play by a batsman/bowler is used for testing, and the rest of the data is used for training purposes. CA is applied to the training data set, and a training biplot is obtained. Similarly, test data is subjected to CA analysis, and a test biplot is obtained. The biplots obtained from the training phase and test phase are compared using the procrustes test. When there is a similarity in the obtained biplots, it is considered that the CA method is reliable. Table~\ref{tab:intrinsic batsman} presents the sum of squared residuals ($ \Delta_{12}^2 $) for eight batsmen\footnote{Procrustes analysis of all the players can be accessed at \url{https://bit.ly/2QZjx5x}}. The lowest sum of squared residual is 0.09 for batsman Joe Root implying the highest similarity for his training and test data set.

	In both the validations high degree of similarity is observed in terms of the derived rules suggesting the accuracy of the proposed method in mining individual player's strength rules and weakness rules. The data and results generated during the validation process are available in the shared repository.
	
	\section{Discussion}\label{sec:D}
	In addition to the batsman's response (strength and weakness), it is important to note the runs scored (\textit{outcome}) on each delivery by the batsman, where the ball is being hit by the batsman (\textit{shot area}) and batsman's \textit{footwork}. From the confrontation matrix of batsman Smith, additional visualizations (biplots) are obtained - (i) \textit{outcome} (refer to Fig.~\ref{fig:InnerA1}), (ii) \textit{shotarea} (refer to Fig.~\ref{fig:InnerB1}), and (i) \textit{footwork} (refer to Fig.~\ref{fig:InnerC1}). 	
	From Fig.~\ref{fig:InnerA1}, the following rules are obtained from Smith's outcome: 
	(i) ``Steve Smith scores high runs on deliveries that are either short-pitched or full length", 
	(ii) ``Steve Smith struggles to score runs and tends to get out on deliveries that are either bowled outside the off stump or moving away". 	
	From Fig.~\ref{fig:InnerB1}, the following rules are obtained from Smith's shotarea: 
	(i) ``Steve Smith plays shots to square leg area on moving in deliveries", 
	(ii) ``Steve Smith plays shots to third man area on moving away deliveries".	
	From Fig.~\ref{fig:InnerC1}, the following rules are obtained from Smith's footwork: 
	(i) ``Steve Smith plays in backfoot on moving in deliveries", 
	(ii) ``Steve Smith plays in frontfoot on deliveries that are either bowled outside the off stump or on spin deliveries". We have shared the biplots (outcome, response, footwork, and shot area) for each player at: \url{https://bit.ly/2SwCQ5r}.

\begin{figure}[!htbp]
		\begin{center}
			\centering
			\subfloat[Subfigure 1 list of figures text][Strength Word Cloud.]{
				\includegraphics[scale=0.46]{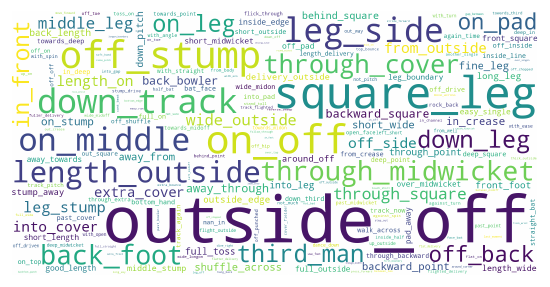}
				\label{fig:subfig1d}} 
			\subfloat[Subfigure 2 list of figures text][Weakness Word Cloud.]{
				\includegraphics[scale=0.46]{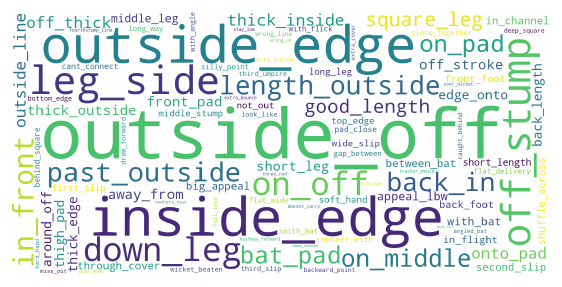}
				\label{fig:subfig2d}}
			\caption{Word Clouds for Batsman Steve Smith.}
			\label{fig:wordclouds}
		\end{center}
	\end{figure}

	We have also investigated the utility of the traditional text visualization techniques in answering R1 and R2. One of the popular text visualization techniques is the word cloud~\cite{TagCrowd}. Word clouds display the relative word frequency or the importance of a word by its font size. Strength (or weakness) word cloud is obtained using the text commentaries which contains the bigrams related to \textit{attacked} (or \textit{beaten}) feature. The strength word cloud for batsman Steve Smith is given in Fig.~\ref{fig:subfig1d}. The most frequent bigram in the word cloud is \textit{outside off}. This is interpreted as a strength rules for Steve Smith i.e. Steve Smith attacks the deliveries that are bowled on \textit{outside off stump}. In a similar fashion, the weakness word cloud for batsman Steve Smith in Fig.~\ref{fig:subfig2d}. 	
	The most frequent bigrams in the weakness word cloud is also \textit{outside off}, i.e. the weakness rule is that Steve Smith gets beaten on the deliveries that are bowled on \textit{outside off stump}. This is a contradiction as the strength rule and weakness rule for Steve Smith can not be identical. Further, in the game of cricket, bowlers frequently deliver outside off stump ball and is a well known fact. This makes the constructed rule trivial and the confidence on such rules will be low in practice. However, in our method even though \textit{outside off} is most frequent in the text commentary, it was not captured in the constructed rules. Contradictions will not be obtained in the proposed method. In addition, a rule comprise of two distinct features namely, batting and bowling. In word cloud based representations, a word can be a batting feature \textit{alone} or a bowling feature \textit{alone}. Therefore the relationship between batting and bowling features cannot be captured using word clouds. Word clouds and all the visualizations which centre around word clouds cannot be used for strength rule or weakness rule construction.  
	
	\section{Conclusion}\label{sec:C} Sports visualization predominantly centers around box-score data and tracking data. For the first time, the utility of the unstructured data is demonstrated. The analysis is focused on advanced computation namely strength rule and weakness rule identification. Visualization of the  obtained rules for each player through biplots is presented. In this work, CA is shown to be a suitable method for computation of such tasks. The constructed rules are validated  using expert analysis and statistical method. The visualizations will be helpful for analysts, coaches and team management in building game strategies.
\pagebreak

    \begin{IEEEbiography}[{\includegraphics[width=1in,height=1.25in,clip,keepaspectratio]{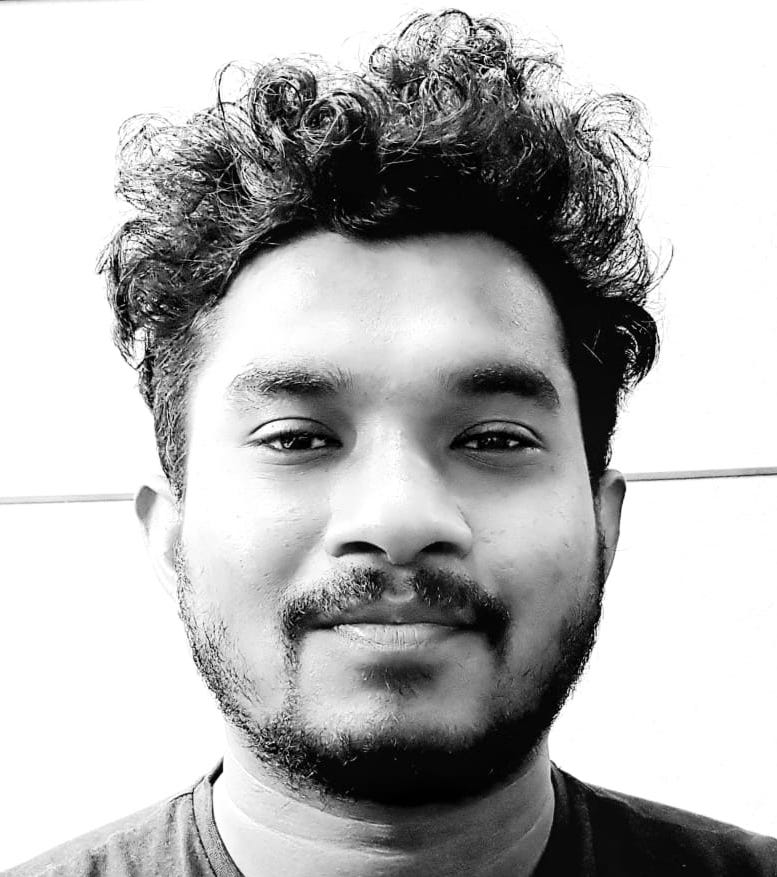}}]{Swarup Ranjan Behera} received the Ph.D. degree in computer science and engineering from Indian Institute of Technology Guwahati, India. He is working as a Senior Research Scientist at NLP-Speech team, AICoE, Jio. His research interests include - natural language processing, audio processing, computer vision, and sports AI.
    \end{IEEEbiography}

	\begin{IEEEbiography}[{\includegraphics[width=1in,height=1.25in,clip,keepaspectratio]{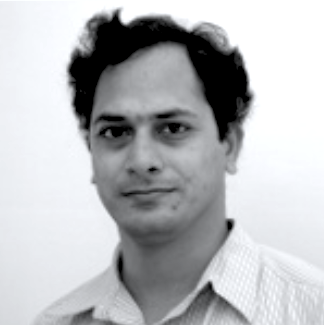}}]{Vijaya V Saradhi} received the Ph.D. degree in computer science and engineering from Indian Institute of Technology Kanpur, India. He is an associate professor with Indian Institute of Technology Guwahati, India. His research interests include - machine learning, kernel methods, data mining, and their applications. 
    \end{IEEEbiography}
  	
\end{document}